# High Performance Hybrid Two Layer Router Architecture for FPGAs Using Network-On- Chip

P.Ezhumalai[1] Dr. C.Arun[2] S.Manojkumar[3] Dr.P.Sakthivel[4] Dr.D.Sridharan[5]

[1&3]Department of Computer Science and Engineering
Sri Venkateswara College of Engineering, Pennalur, Sriperumbudure-602105, Chennai, Tamilnadu, India

[2]Department of Electronics & Communication Engineering
Rajalakshmi Engineering College Thandalam, Chennai - 602 105, Tamilnadu, India

[4&5]Department of Electronics & Communication Engineering
College of Engineering, Guindy Anna University, Chennai, Tamilnadu

*Abstract*— Networks-on-Chip is a recent solution paradigm adopted to increase the performance of Multi-core designs. The key idea is to interconnect various computation modules (IP cores) in a network fashion and transport packets simultaneously across them, thereby gaining performance. In addition to improving performance by having multiple packets in flight, NoCs also present a host of other advantages including scalability, power efficiency, and component re-use through modular design. This work focuses on design and development of high performance communication architectures for FPGAs using NoCs

Once completely developed, the above methodology could be used to augment the current FPGA design flow for implementing multi-core SoC applications. We design and implement an NoC framework for FPGAs, Multi-Clock On-Chip Network for Reconfigurable Systems (MoCReS).We propose a novel micro-architecture for a hybrid two-layer router that supports both packet-switched communications, across its local and directional ports, as well as, time multiplexed circuit-switched communications among the multiple IP cores directly connected to it. Results from place and route VHDL models of the advanced router architecture show an average improvement of 20.4% in NoC bandwidth (maximum of 24% compared to a traditional NoC). We parameterize the hybrid router model over the number of ports, channel width and bRAM depth and develop a library of network components (MoClib Library). For your paper to be published in the conference proceedings, you must use this document as both an instruction set and as a template into which you can type your own text. If your paper does not conform to the required format, you will be asked to fix it.

*Keywords:* Core Based Design, FPGA, Network on Chip (NoC), On Chip Communication, MoCReS, System on Chip (SoC),

## I. INTRODUCTION

The two main concerns with NoC designs that are strictly packet-switched are the control and serialization overhead involved in transferring data between IP cores that are placed close to each other in the FPGA. In order to ensure high throughput between these cores, we advocate time-multiplexed circuit-switched connections. In addition to this mode of transfer, the router also preserves the online nature of communication between farther cores through the packet-switched layer. The area efficient MoCReS architecture is modified to support both the above mentioned layers of operation. The design goals and issues involved in the hybrid two-layer architecture are presented in this paper. We also develop a SystemC model of our router for both functionally verifying the design as well as to vary its specifications and obtain the performance results rapidly through simulation. We present the results and analysis of the novel router architecture in this paper.

We target our proposed NoC framework for reconfigurable computing platforms and therefore we restrict our discussions in this section primarily to existing FPGA based NoCs. NoCs were introduced into the FPGA domain mainly to simplify tile-based reconfiguration [1] [2], and its potential as effective communication architecture is largely unexplored [7].
Research in [5] [6] address the capabilities of FPGAs to support NoC based multi-processor applications. Hilton et al. [4] incorporate flexibility into their design for FPGA based circuit-switched NoCs. However, their strictly circuit-switched router suffers from signal integrity and path reservation issues which we overcome in our design. SoC BUS [8] proposes a circuit-switched router with a packet based setup. Here, control packets are responsible for setting up strict circuit-switched connections, which is different from our two-layer approach. Research in [9] [4] [3] also present FPGA based NoCs. The above designs ignore implementation level area-performance trade-offs while proposing the architecture, thereby limiting to a system-level performance analysis. To the best of our knowledge, this is the first work to propose an FPGA-suitable hybrid router architecture integrated with an automatic topology synthesis framework that satisfies the bandwidth requirements of an application while optimizing its area overhead

## II. Motivation

Packet-switching performs online scheduling by dynamically negotiating communication between the cores. An alternate technique, namely circuit-switching offers high through-put dedicated connections to overcome the performance drawbacks in packet-switching by scheduling time-multiplexed communication across the cores. Even though this static scheduling requires all the communication patterns to be known before hand, it can pro-vide a very high throughput with marginal area overhead (for storing schedules). We propose a modified router architecture which interfaces multiple IP cores to the router and supports packet-switching for inter router transfers and time-multiplexed circuit-switching for IP cores connected to the same router. This technique also eliminates the latency in req/grant protocol, serialization and control overheads for data transfers between cores placed close to each other in FPGAs and mapped to the same router.

### A. Packetization and Control Overheads

In this section, we quantify the overheads associated with the existing baseline approach (MoCReS). Control and Packetization are the two main overheads associated with the MoCReS framework.

*Control Overhead:* In MoCReS, connections between various ports are established through a req/grant protocol which involves round-robin arbitration in the case of common ports requests (conflicts). We see that it takes at least 6 cycles for the data at the input port to appear at the output of a router (as input to the downstream router/local IP). This setup latency is a fixed overhead in addition to the delays due to network congestion.

*Packetization Overhead:* Due to the nature of interconnection network, the channel width between ports/routers are limited to a fixed size (8 bits in MoCReS, baseline version). Due to this fixed channel width, the communication data that is to be sent over the network must be quantized into flits. Variable number of flits constitutes a packet. If F is the number of flits in a packet and b is the channel width, then F/b is the serialization latency associated with the communication.

## III. Architecture Description

In this section, we first present the modified router micro-architecture, followed by its architectural advantages and design issues involved. The network topologies along with the Flow controls for the packet-switched layer are kept the same as presented in this paper [10]. **Network Topology:** Mesh networks have minimum area overhead (reduced long lines) [5] [10], low power consumption and map well to the underlying routing structure of FPGAs. Hence, we choose a mesh topology to optimize logic and routing in FPGAs, and to provide sufficient resources for the IP cores.

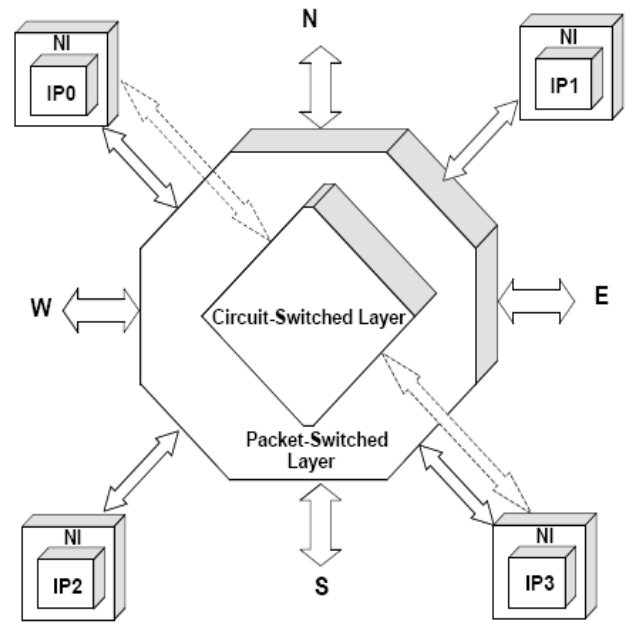

Figure 1 : Hybrid Two-Layer Router Architecture

*Flow Control:* Our router supports multi-clock virtual cut-through flow control with a deadlock-free XY routing. The switch complexity involved in the above choice is more suitable for a light-weight implementation [10].

### A. Cross-Point Matrix

**Architecture Modifications:** The modified switch is comprised of two layers of operation: a high throughput time-multiplexed circuit-switched layer (C-layer) and a multi-clock packet-switched layer (P-layer). Variable number of IP cores connected to the switch participates in the C-layer, thereby achieving guaranteed throughput and more predictable latencies between IP cores placed close to each other in the FPGA.

Figure 1 presents the novel two-layer hybrid router architecture. This modified router has four local IP ports, in addition to the four directional ports. Further, in this case two of the four local IPs (IP 0, IP 3) are participating in the time-multiplexed circuit-switched layer. Using the packet-switched layer, all the four IPs can communicate to the neighboring routers through the directional ports. The cross-point matrix is multiplexer based, as opposed to providing connections for each virtual channel. The following are the design issues involved with the cross-point.

*Packet-Switched Cross-Point:* In the packet-switched layer, the directional input ports (N, E, and S, W) are multiplexed to every local port. Therefore cross-point connections are introduced to support these additional local ports. However, all the connections between the local ports in this layer are removed, as they are connected in the circuit- switched layer. The ports connected through the C-Layer (IP 0, IP 3) cannot participate in the P-Layer to transfer data between themselves. This translates into gain in area which we utilize to increase the bandwidth available.

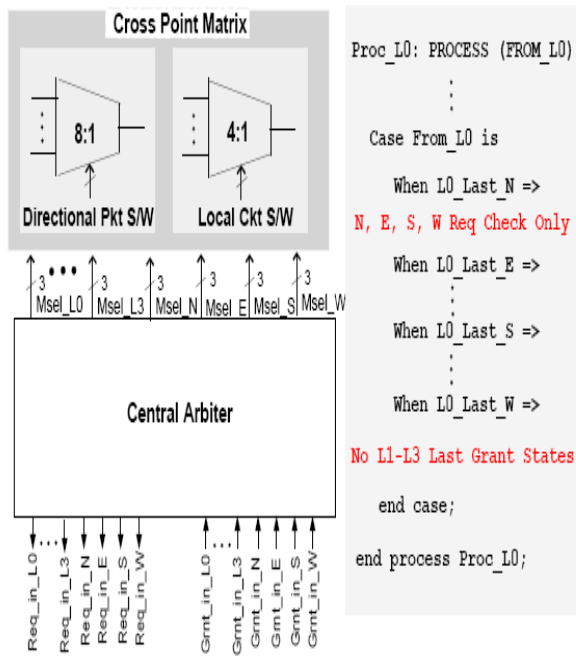

Figure 2: Modified Central Arbiter Model

*Circuit-Switched Cross-Point:* Let $L_i$ is the total number of local IPs and $P_i$ is the number of ports participating in the circuit-switched layer. The bus width of this cross-point is currently set to 32 bits in order to support a very high bandwidth. Further, this cross-point can handle a maximum of $P_i$ high throughput parallel connections. The scheduling memory configures this cross-point during various time slots.

*Router Channel Widths:* Due to high throughput requirement between the cores participating in the circuit-switched layer, we set the channel width to 32 bits (corresponding to the data width of micro blaze soft processor). In the packet-switched layer, we retain the bus width of MoCReS (8 bits/channel). However, choice of an appropriate channel width is a trade off between resources available and bandwidth required.

### B. Central Arbiter

The Central Arbiter is responsible for configuring the simultaneous connections by setting the cross-point in the P-Layer. We run parallel FSMs to ensure that no queuing takes place between requests. As long as the participating IPs request mutually exclusive ports, the connections happen parallel. In case of queuing/conflicts, the arbitration is performed through the round robin approach. The IPs that participates in the C-Layer will not need arbitration between them in the P-Layer. We perform state reduction in the FSMs corresponding to those inter-local port connections .i.e in correspondence with the inter local IP connections that are removed (Section 3.1) in the packet-switched layer. The Central Arbiter is also customized to not support states for these connections. The simplicity of round-robin arbitration coupled with the above state reduction translates into significant area savings. Figure 2 shows the modified central arbiter model.

### C. NI Design

The network interface arbitrates the choice of packet/circuit switched layer and is also responsible for supporting variable size packets. Mode Switching: Upon receiving the target IP co-ordinates, it triggers the mode signal to decide if the packet will be decoded to leave the router or the cross point is triggered in circuit switch mode.

*Variable Packet Sizes:* In during packet-switched transfer, the network interface is also responsible for encoding the header with:

1. Packet Size (As a fraction of bRAM depth)

2. X co-ordinate of destination IP

3. Y co-ordinate of destination IP

The packets transferred through the network can be broadly classified as control (lesser number of flits) or data. Therefore, the packets will be of varied sizes. The NI encodes the packet size as a fraction of the total bRAM depth along with the header. This novelty improves buffer utilization, thereby increasing the performance of the NoC.

### D. Design Parameters

In order to quickly explore the NoC design space, we have parameterized the structural

VHDL model of our router for:

1. Total number of ports
2. Channel width
3. Virtual Channels/port
4. Number of ports participating in the C-Layer

By varying the above parameters, we develop a component library, MoC lib which we use to characterize variants of the router for area and operating frequency.

## IV. Architectural Advantages

**Bandwidth Increase:** Bandwidth available in a switch is the product of the number of ports, operating frequency and channel width. The C-layer has minimum logic overhead with no buffering and can operate at a clock rate significantly higher than the P-layer. Furthermore, increasing the number of ports also scales the available band-width in a switch. Moreover, the absence of control/serialization overheads (req/grant) also increases the throughput.

**Power Savings:** The amount of logic required for the NoC reduces with router count, thereby saving static power. Further, with increasing number of ports within a router, the average packet latency is also reduced [9]. Therefore dynamic power drops considerably with reduction of router hops.

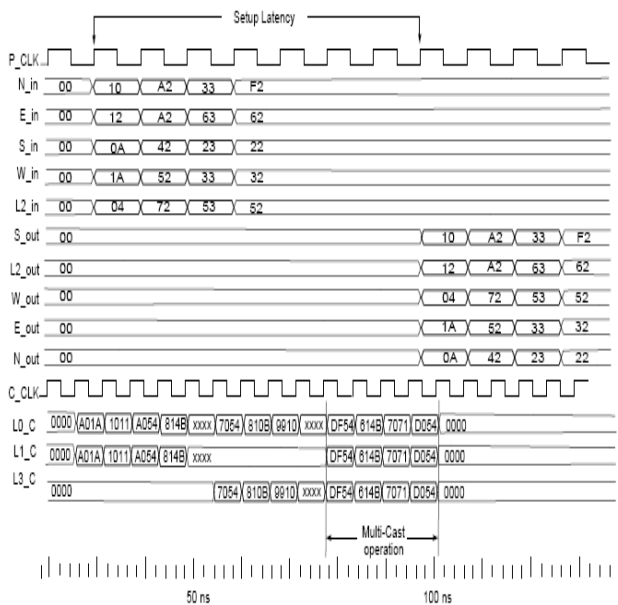

Figure 3: SystemC Simulation

**Guaranteed Throughput:** The time-multiplexed nature of the C-Layer scheduling provides good Quality of Service (QoS) to the application, particularly, between cores placed close to each other. Otherwise, the NoC would have to support area expensive QoS protocols to ensure the required bandwidth.

**Inherent Multi-Cast Capability:** The cross-point in the C-layer can be configured simultaneously for a multi-cast (one to many destinations) operation among IPs connected to the same router without any penalty in performance. Further, this capability also optimizes the area required for storing the schedules (with fewer bits required to encode the configuration data of the circuit-switched network).

## V. System-Level Router Model

With increasing design complexities, there is a need for rapid design space exploration that makes use of a set of specifications. We model our NoC router framework using SystemC. By doing so, we functionally verify the model as well as setup a platform to estimate the advantages of this architecture over the baseline approach.

SystemC is a description language that abstracts the computation elements of a design by behaviours (or processes) and simplifies the communication between the cores using transaction level modelling. The framework has a set of library routines and macros implemented using C++. The behaviour of the hardware to be modelled is captured by simulating concurrent processes coded in C++.

SystemC Tool Flow: Every component in the router is modelled in C++ as a process. This .cpp file can be compiled and executed with the SystemC engine that is written in C++. We use the open source SystemC version 2.1 to compile our router design. The set of .cpp files are first compiled with the appropriate command options. Then, an executable is created to run the tool flow. We dump out the Value Change Dump (VCD) File from the engine.

The .VCD file of the router model can be used as follows:

• Applied to standard simulation tool for verifying the functionality of the model by viewing the waveform

• Estimate preliminary power consumed by the implementation on FPGAs, by using

Xpower and the architecture information (Virtex-4)

## VI. Synthesis Results

In this section we present the Area/Synthesis results for our modified router implemented on Xilinx Virtex 4 [11]. The additional bandwidth offered by the proposed router comes with an increase in switch complexity. The amount of FPGA logic and routing resources consumed by the router instance depends on its complexity

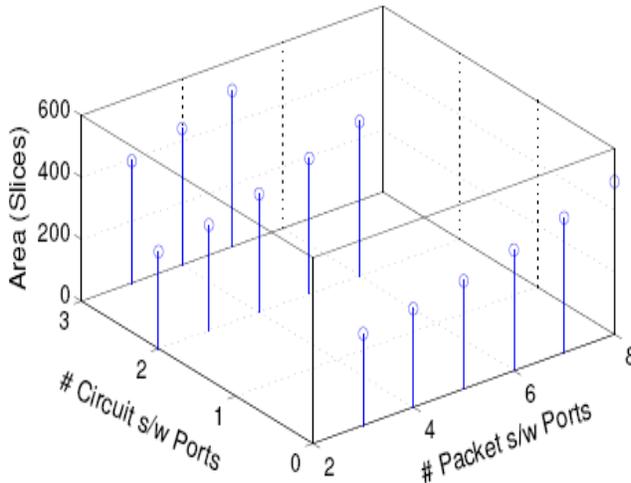

Figure 4: Design Parameters Vs Area

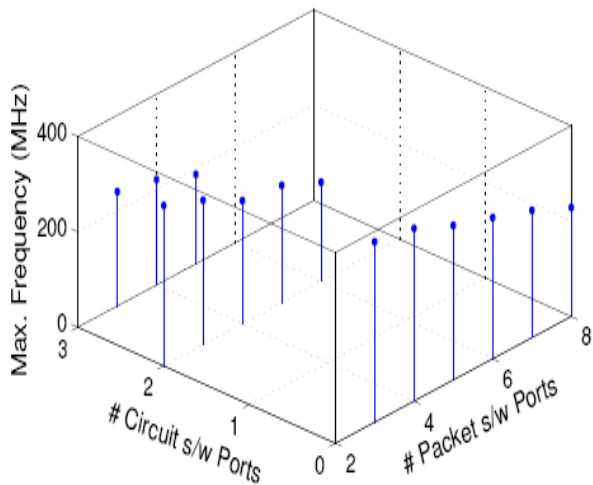

Figure 5: Design Parameters Vs Frequency

Figure 4 presents this variation in switch area with the number of ports (C & P-Layer) it supports.

Further, the operating frequency of the router instances vary greatly due to different critical path lengths. Also, with increasing number of ports participating in the circuit-switched layer, the routing resources deplete rapidly (due to increased channel widths). This degradation in Performance in turn affects the bandwidth the switch can offer. Figure 5 presents the variation in switch operating frequency with the number of ports in both layers. The above area and frequency estimates are obtained by varying the parameters in the VHDL model of the router and by implementing them on the target device

Table 1: Scaling of Area and Frequency with No. of C-Layer Ports

| MoClib Component | Area (Slices) | Frequency (MHz) |
|---|---|---|
| MC (4,2,2) | 314 | 336 |
| MC (5,3,2) | 326 | 318 |
| MC (5,2,3) | 341 | 303 |
| MC (6,3,3) | 394 | 240 |
| MC (6,2,4) | 382 | 258 |
| MC (7,3,4) | 440 | 221 |

Furthermore, to perform automatic topology synthesis, we estimate the increase/decrease in switch area with exclusive variations in number of P-Layer ports and C-Layer ports independently. When NoC area is in the cost function, the above data will aid rapid design space exploration. Tables 1 and 2 present the scaling of area & frequency with increasing C-Layer and P-Layer ports respectively. In the tables, MC(x, y, z) denote an instance of the MoC lib library, where y is the total number of C-Layer ports, z is the total number of P-Layer ports and x is the sum of the two (total number of ports). Table 2 presents the scaling of area and frequency only with respect to the P-Layer ports and therefore they can be considered as variations of the MoCReS baseline router.

Table 2: Scaling of Area and Frequency with No. of P-Layer Ports

| MoClib Component | Area (Slices) | Frequency (MHz) |
|---|---|---|
| MC (3,0,3) | 296 | 378 |
| MC (4,0,4) | 318 | 362 |
| MC (5,0,5) | 349 | 324 |
| MC (6,0,6) | 390 | 296 |
| MC (7,0,7) | 435 | 267 |
| MC (8,0,8) | 493 | 229 |

VII. Results: Performance Improvement

*Area vs Average Available Bandwidth/Port:* The baseline version in this com- parson is MoCReS with 1VC+MC. The area (in slices) of the switch increases with the number of ports it supports. We measure the area values for increasing number of ports (packet-switched) in the baseline version. For similar area values, when the alternate hybrid router is used, there is an increase in available bandwidth per port. This band- width increase associated with the hybrid router architecture is compared in this section with the baseline approach. For equivalent area overheads (in slices) on a similar FPGA, Figure 6 presents the bandwidth capacity (in MB/s) of the NoC (per port) for both approaches. In spite of a rapid degradation in operating frequency (with increase in circuit-switched ports), there is a significant bandwidth gain using the hybrid two-layer approach. For the area window utilized in our library of routers, there is an average 20.4% gain in bandwidth (maximum of 24%) offered by our NoC. This gain in performance is due to supporting a high throughput circuit-switched layer with a marginal area overhead.

### A. Design Issues

Even though it appears intuitively that an increase in number of ports in the C-layer gives performance benefits without any area overhead, there are certain design issues that can potentially limit the performance due to increase in switch complexity.

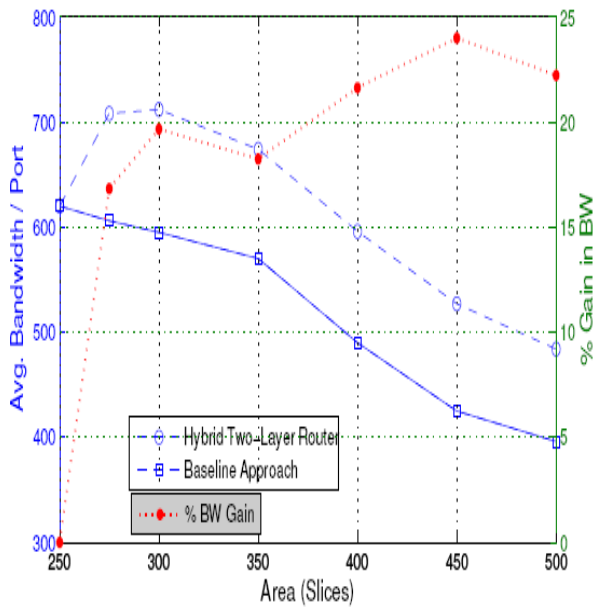

*Figure 6: Area (Slices) Vs Avg. Bandwidth / Port*

Operating frequency Vs Switch Complexity: There is a depletion of critical re- sources associated with an increase in switch complexity (number of ports, bus width). As a result, the operating frequency of the switch degrades which in turn affects the bandwidth offered by the router. For the NoC paradigm to efficiently be an alternative to the bus-based architecture, the performance design parameters must be chosen carefully so that it is possible to operate the routers at the highest possible frequency. Switch Power vs Link Power: By increasing the number of ports, we can reduce the average hop count [9], i.e we minimize the routers and links. This translates into a reduction in power consumed by the links, but an increase in power consumed by the switches. Beyond a cut-off, the increase in switch power can potentially overshadow the gain in link power; thereby it can increase the power/flit ratio

Explosion of Schedule Memory: With increasing number of C-layer ports, the schedule memory also scales linearly. The schedule memory, expressed in number of LUTs is a function of number of schedule cycles and C-layer ports present. If C is the number of ports participating in the C-Layer, then $\log_2 C$ is the number of configuration bits required per cycle.

*Clock Signal Integrity:* Operation of the C-layer ports requires the participating IP cores to be synchronous, as there is no buffering done, as opposed to packet-switch where multi-clock FIFOs separate the clock domains. Increasing the number of C-layer ports could potentially increase the distance between the connected IP cores. In this case, the signal integrity acts as a limitation to the number of C-layer ports, and reduces the clock rate.

It can be seen that all of the above factors limit the amount of performance gain that can be achieved using our hybrid approach. This trade-off between performance, area and port count merits a balance and requires an application-suitable tuning of the NoC topology.

### VIII. Conclusions

To address the bandwidth limitations of MoCReS, we extend the design by developing hybrid two-layer router architecture. The novel design of the network component supports high throughput time-multiplexed circuit-switched connections between IPs interfaced to the same router, in addition to the packet-switched communication layer. Various instances of the NoC components are characterized for area and performance in the form of an MoC lib NoC component library. The advanced router architecture achieves an average improvement of 20.4% in NoC bandwidth (maximum of 24% compared to a traditional NoC).

# REFERENCE

[1] Theodore Marescaux et al. Interconnectin Networks Enable Fine-Grain Multi-Tasking on FPGAs. In FPL'2002, pages 795–805, 2002.

[2] A.Kumar et al. An FPGA Design Flow for Reconfigurable Network-Based Multi-Processor Systems on Chip. In DATE'07, 2007.

[3] N.Kapre. Packet-Switched On-Chip FPGA Overlay Networks. MS thesis, California institute of technology. 2006.

[4] Clint Hilton and Brent Nelson. PNoC: a flexible circuit-switched NoC for FPGA based systems. In IEEE Proc. Computers and Digital Techniques, 2006.

[5] Manuel Saldaa, Lesley Shannon, and Paul Chow. The Routability of Multiprocessor Network Topologies in FPGAs. In SLIP'06, pages 49–56, 2006.

[6] T.A Bartic et. Al. Topology Adaptive Network-on-Chip Design and Implementation. In Computer and Digital Techniques, IEE Proceedings, pages 467–472, 2005.

[7] T.S.T. Mak et.al. On-FPGA Communication Architectures and Design Factors. In FPL'06, 2006.

[8] D. Wiklund and L.Dake. SoC BUS: switched network on chip for hard real time embedded systems. In Parallel and Distributed Processing Symposium, 2003.

[9] Balasubramanian Sethuraman and Ranga Vemuri. OptiMap: a tool for automated generation of NoC architectures using multi-port routers for FPGAs. In Design, Automation and Test in Europe, 2006. DATE '06, 2006.

[10] A.Janarthanan et.al. MoCReS: an Area-Efficient Multi Clock On-Chip Network for Reconfigurable Systems. In IEEE Computer Society ISVLSI'07, 2007.

[11] Xilinx Inc. http://www.xilinx.com.

## AUTHORS PROFILE

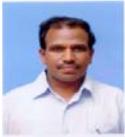
Ezhumalai Periyathambi received the B.E degree in Computer Science and engineering from Madras University, Chennai, India in 1992 and Master Technology (M.Tech.,) in computer science and Engineering from JNT University, Hyderabad, India in 2006. He is currently working towards the Ph.D degree in Department of Information and Communication, Anna University, Chennai, India. He is working as assistant Professor in the Department of Computer Science and Engineering, Sri Venkateswara College of Engineering, sriperumbudur, Chennai, Tamilnadu, India. His research in reconfigurable architecture, Multi-Core Technology CAD – Algorithms for VLSI Architecture. Theoretical Computer Science. and mobile computing.

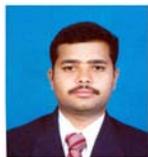
Arun Chokkalingam received the B.E degree in electronics and communication engineering from Bharathidasan University, Trichy, India in 2002 and the M.E degree from Anna University, Chennai, India 2004 and Doctorate in VLSI design at Anna University, Chennai TN, India in the year 2009. He is currently working towards the Ph.D degree in Department of Information and Communication, Anna University, Chennai, India. Since 2004 he has been an Lecturer in the Department of Information Technology, Sri Venkateswara College of Engineering, Chennai, Tamilnadu, India.His research in error correcting codes addresses effectively decoding algorithm and VLSI Architecture. His research interest including digital communication, coding theory, modulation and mobile communication.

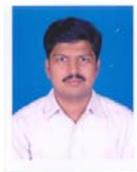
S.Manoj kumar received the BE degree in Computer Science and Engineering from Bharadhidasan University, India in 2002 and Master of Engineering (ME) in Computer Science & Engineering from Anna University, Chennai, India in 2008. He is working as faculty in the Department of Computer Science and Engineering, Sri Venkateswara College of Engineering, sriperumbudur, Chennai, Tamilnadu, India. His research in reconfigurable architecture, Networking and mobile computing.